\documentclass[12pt]{article}

\usepackage[margin=1in]{geometry}
\usepackage{amsfonts}
\usepackage{amsmath, amssymb}
\usepackage{amsthm}
\usepackage{setspace}
\usepackage{graphicx}
\usepackage{booktabs}
\usepackage{caption}
\usepackage{newpxtext}
\usepackage{newpxmath}
\usepackage[sorting=none,citestyle=numeric-comp]{biblatex}
\usepackage{hyperref}
\usepackage{authblk}
\usepackage{csquotes}
\usepackage{dsfont}

\definecolor{mao}{RGB}{202, 53, 27}
\definecolor{grape}{RGB}{63, 37, 110}
\definecolor{ocean}{RGB}{29, 99, 174}
\hypersetup{
    colorlinks=true,
    linkcolor=grape,
    filecolor=grape,      
    urlcolor=grape,
    citecolor=grape
}

\newcommand{\g}{\text{g}}
\newcommand{\id}{\mathds{1}}

\def\id{\mathds{1}}

\newcommand{\pureket}[1]{|{#1}\rangle\!\langle{#1}|}

\title{The quantum-gravitational imitation game}

\author[1]{Kristian Toccacelo}
\affil{\small\textit{Center for Macroscopic Quantum States (bigQ), Department of Physics,
Technical University of Denmark, 2800 Kongens Lyngby, Denmark}}
\affil{\texttt{kristo@dtu.dk}}

\date{\today}

\date{\today}

\begin{document}

\maketitle

\begin{abstract}
\noindent Gravity is the most apparent force in our everyday existence. Yet its fundamental nature remains the most opaque of the known interactions. This gap in our understanding is, in large part, due to the weakness of the gravitational interaction, which makes its empirical probing exceedingly hard. Nevertheless, on the backdrop of rapid advances in quantum technologies, hope has mounted that tests of the quantum nature of gravity could be realized in tabletop experiments. In this essay, we frame these recently proposed tests as \textit{quantum-gravitational imitation games}. In particular, we examine how gravitational interactions among mechanical oscillators enable the teleportation of arbitrary quantum states and how this can inform fundamental tests of gravity.     
\end{abstract}

\begin{center}
\textit{Selected for honorable mention in the Gravity Research Foundation 2026 Awards for Essays on Gravitation.}
\end{center}

\newpage

\subsection*{Introduction}
The question \textit{Is gravity quantum?} is somewhat ambiguous. While the terms of the query can be in principle defined properly, in a strict Popperian sense, such yes/no questions ultimately cannot be answered scientifically~\cite{popper2005logic}. Still, physicists today might, more likely than not\footnote{\doublespacing There seems to be no survey that asks physicists this question exactly. However, we can infer from a survey conducted at the "Black Holes Inside and Out" conference held in Copenhagen in the summer of 2024 \cite{chen2025copenhagensurveyblackholes}. When asked which is the best candidate for a theory of quantum gravity, $9\%$ of respondents believed gravity is not quantum, and $24\%$ had no opinion, meaning at least $67\%$ of respondents believe gravity is quantum.}, reply to the prompt with something along the lines of: "Well, yes, of course!" If we take ChatGPT as a ghostly representation of this average, its response to the query is quite revealing: "Short answer: we strongly suspect gravity is quantum---but we haven’t conclusively proven it yet." But what exactly counts as conclusive proof? We could mean, for instance, that the quantization of the gravitational field is necessary on logical grounds. While there is a rich tradition of arguments in favor of quantizing the gravitational field \cite{dewitt1962definition,eppley1977necessity,mari2016experiments,oppenheim2023time}, this logical necessity must ultimately be grounded in empirical necessity. This gets to the ambiguity of the question posed at the beginning. It would never really be possible to prove that the gravitational field is quantized. No experiment could do that. We can only ever hope to empirically distinguish among different hypotheses of how reality works. The productive question is therefore not \textit{Is gravity quantum?} but rather: \textit{What can a quantum theory of gravity do that a classical theory cannot?}

\subsection*{The quantum-gravitational imitation game}
\begin{figure}[t]
    \centering
    \includegraphics{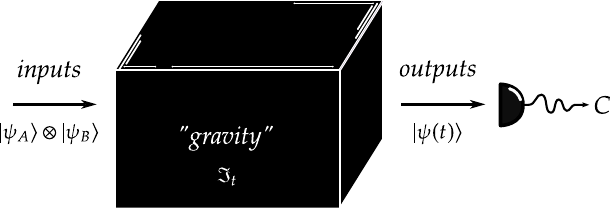}
    \caption{\doublespacing \textit{The quantum-gravitational imitation game.} Alice and Bob feed some input state to a gravitational black box controlled by Isaac. Isaac realizes the gravitational interaction by means of the CPTP map $\mathfrak{I}_t$ on the joint input state $|\psi_A\rangle\otimes|\psi_B\rangle$. The outputs are then analyzed by Charlie, who, without knowing details of $\mathfrak{I}_t$, verifies whether the operations applied by Isaac are classical or quantum. }
    \label{fig:imitation_game}
\end{figure}

Some ambiguity remains in the question as reframed above. There is no agreed-upon, consistent theory of quantum gravity, and what counts as "classical" is itself underspecified. We return to these issues later. For now, let us formalize the question by way of a game we call the \textit{the quantum-gravitational imitation game}\footnote{\doublespacing Taking inspiration from Alan Turing's \textit{imitation game} introduced in his famous 1950 article \textit{Computing machinery and intelligence} \cite{turing2007computing}.} illustrated in Fig.~\ref{fig:imitation_game}. 

The game is played by four parties: Alice ($A$), Bob ($B$),  Isaac ($I$), and Charlie ($C$). Alice and Bob represent two quantum systems of electrically neutral matter distributions with Hilbert spaces $\mathcal{H}_A$ and $\mathcal{H}_B$. At the initial time $t_0=0$, $A$ and $B$ prepare the local pure states $|\psi_A\rangle$ and $|\psi_B\rangle$ so that the joint state of $A$ and $B$ together is the product state $|\psi\rangle=|\psi_A\rangle\otimes|\psi_B\rangle$. They then hand this joint state to Isaac, who represents the gravitational interaction. Isaac implements a completely positive trace-preserving (CPTP) map---the most general transformation consistent with the quantum-mechanical nature of the matter systems---on the joint state over a time interval of duration $t$, during which no other interactions can affect the evolution.  Finally, Isaac hands over the evolved state $\mathfrak{I}_t(\psi)=|\psi(t)\rangle$ to the fourth party, Charlie.  

Importantly, details of the possible implementation of the transformation applied by Isaac are unknown to Charlie. From Charlie's perspective, the gravitational interaction is a black box. The role of Charlie is that of a verifier whose goal is to decide whether the interaction, namely the transformation applied by Isaac, can be realized by means of classical rules. To do so, the verifier issues a "challenge" to the gravitational interaction by looking for a set of observable signatures in the transformed states that are uniquely consistent with the hypothesis that the gravitational interaction is quantum in nature. We refer to the quantum hypothesis as the \textit{null hypothesis}. If such signatures are found, competing \textit{alternative hypotheses} can be excluded. 

Since our motivation is to determine whether gravity needs to be quantized, the null hypothesis need not assume a full theory of quantum gravity. An effective description suffices. We then take as our baseline null hypothesis gravity perturbatively quantized around the flat Minkowski metric. The alternative, classical hypothesis is motivated by two assumptions. First, the gravitational field is a classical field configuration specified by the metric tensor $g_{\mu\nu}^{(j)}(x)$ that can occur with probability $P^{(j)}$. Second, two systems with certain matter distributions outside each other's support (that is, spacelike separated) can couple only to their local fields. The set of CPTP transformations on the states prepared by Alice and Bob that are consistent with these assumptions is the set of local operations assisted by classical communication (LOCC).

The game is thus designed to rule out the entire class of LOCC-compatible models of gravity, in direct analogy with Bell tests. In fact, to the point made earlier, the experimental violation of a Bell inequality does not "prove" the fundamental "quantumness" of the world, but rather, excludes a whole class of local hidden variable models in contradiction with the violation. Let us now illustrate a concrete example of such a test.

\subsection*{Gravitational teleportation}
\begin{figure}[t]
    \centering
    \includegraphics{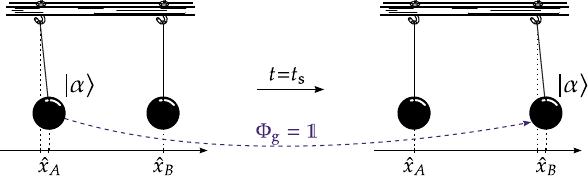}
    \caption{\doublespacing \textit{Gravitational teleportation.} The gravitational interaction between two harmonic oscillators results in the beamsplitter dynamics in Eq.~\eqref{eq:bs}, which at time $t_\mathrm{s}$ implements an effective teleportation channel between Alice's input state (here the coherent state $|\alpha\rangle$) and Bob's output.}
    \label{fig:gravitational_teleportation}
\end{figure}

Suppose Alice and Bob each control a mechanical oscillator of mass $m$ and frequency $\omega$, separated by an equilibrium distance $d$. In the non-relativistic, weak-field limit, the leading-order gravitational coupling between their position fluctuations takes the form of a beamsplitter interaction,
\begin{align}
\label{eq:coupled_oscill}
    \hat{H}_\mathrm{int} = \hbar\,\gamma_\mathrm{g}
    \left(\hat{a}\hat{b}^\dagger + \hat{a}^\dagger\hat{b}\right),\,\,\text{where} \,\,\, 
    \gamma_\mathrm{g} = \frac{Gm}{\omega d^3},
\end{align}
obtained by expanding the Newtonian potential to leading order in the displacements and by applying the rotating wave approximation, valid since $\gamma_\mathrm{g} \ll \omega$ for all realistic mechanical systems.  In the Heisenberg picture, the evolution generated by Eq.~\eqref{eq:coupled_oscill} takes a particularly simple form, described by the beamsplitter transformations
\begin{align}
\begin{split}
\label{eq:bs}
   \hat{a}(t) &= e^{-i\omega t} \left[\hat{a}(0) \cos(\gamma_\mathrm{g} t)- i \hat{b}(0) \sin(\gamma_\mathrm{g} t)\right], \\ \hat{b}(t) &= e^{-i\omega t} \left[\hat{b}(0) \cos(\gamma_\mathrm{g} t)- i \hat{a}(0) \sin(\gamma_\mathrm{g} t)\right].
\end{split}
\end{align}
At the time $t_\mathrm{s} = \pi/(2\gamma_\mathrm{g})$, the two modes are exchanged up to local phases. As a result, any initial product state undergoes a SWAP operation, meaning that the unitary evolution $\hat{U}_t=\exp\big(-i\hat{H}_\mathrm{int}t/\hbar\big)$ acts as
\begin{align}
\hat{U}_{t=t_\mathrm{s}}\bigl(|\psi\rangle_A\otimes|0\rangle_B\bigr) = |0\rangle_A\otimes|\psi\rangle_B.
\end{align}
In other words, the gravitational channel $\Phi_\mathrm{g}:\mathcal{H}_A\to\mathcal{H}_B$ that maps Alice's state at $t=0$ to Bob's state at $t=t_\mathrm{s}$ is precisely the identity channel, $\Phi_\mathrm{g}=\id$ (see Fig.~\ref{fig:gravitational_teleportation}). Gravity has implemented a teleportation channel! This constitutes our null hypothesis. The question is then whether this channel can be mimicked by classical means. This is the quantum-gravitational imitation game. 

Suppose that Alice has access to the ensemble of states $\mathcal{E}_A=\{p_x,\psi_x\}_x$, where $p_x$ are probabilities, and $\psi_x=\pureket{\psi_x}$ are pure states. At each round of the game, Alice draws a state from $\mathcal{E}_A$ at random, while Bob prepares the vacuum $|0\rangle_B$. To decide whether Isaac implemented the quantum gravitational transformation $\Phi_\g$ or a classical LOCC channel $\Lambda$, the verifier can simply check whether the measured output state on Bob's side matches Alice's input state. Quantitatively, this can be achieved by determining the overlap $F=\langle\psi_x|\mathfrak{I}(\psi_x)|\psi_x\rangle$, which over many rounds of the game leads to the average fidelity
\begin{align}
\overline{F}=\sum_xp_x\langle\psi_x|\mathfrak{I}(\psi_x)|\psi_x\rangle.
\end{align}
This quantity equals unity when the null hypothesis holds. Now, suppose Isaac were secretly a classical agent, armed with knowledge of which test the verifier is carrying out. His optimal strategy for deceiving Charlie and passing the quantum imitation game is to maximise the average fidelity, by whatever means. This defines the classical threshold
\begin{align}
\label{eq:unitary_locc_simulation_fidelity}
    F_{c\ell}\equiv\sup_{\Lambda\in\mathrm{LOCC}}\sum_xp_x\langle\psi_x|\Lambda(\psi_x)|\psi_x\rangle,
\end{align}
also known as the classical LOCC bound \cite{lami2024testing}. 
Charlie is thus able to distinguish the null from alternative hypotheses whenever the classical threshold is bounded away from unity, that is, $F_{c\ell}<1$. If the verifier observes an output fidelity exceeding $F_{c\ell}$, the classical hypothesis is rejected. This is a well-known quantum information-theoretic result stating that faithful teleportation of unknown quantum states is possible only with quantum resources. In general, the optimization over all LOCC channels above is not mathematically tractable. However, we need not determine an exact bound. An upper bound based on a set approximating LOCC suffices. In fact, given $F_{c\ell}\leq\mathcal{F}\leq1 $, witnessing a fidelity larger than $\mathcal{F}$ is sufficient to rule out the alternative hypothesis. Such inequalities are called \textit{LOCC inequalities} \cite{lami2024testing}, in analogy with Bell inequalities. 

A particularly simple class of oscillator states Alice might have access to are coherent states. In particular, if the initial state on $A$ is picked at random from the ensemble  $\mathcal{E}_A=\{p_\alpha,|\alpha\rangle\}$, with Gaussian prior $p_\alpha=\lambda e^{-\lambda|\alpha|^2}/\pi$, the best average teleportation fidelity achievable by a classical channel $\Lambda$ is \cite{hammerer2005quantum}
\begin{align}
    F_{c\ell}=\frac{1+\lambda}{2+\lambda}.
\end{align}
The test is inconclusive in the limit $\lambda\to\infty$, where the bound approaches unity. Conversely, the largest deviation between null and alternative hypotheses is reached when $\lambda\to0$, for which the threshold approaches $1/2$. For finite $\lambda$, $F_{c\ell}$ is bounded away from unity. 

Naturally, the question arises of whether such a test can be implemented in practice. In general, due to the weakness of gravitational interactions, it would be difficult to observe complete teleportation via gravitational interactions as described here. Nonetheless, it is possible to prove classical thresholds for the intermediate "partial teleportation dynamics" at times $t<t_\mathrm{s}$, for which the bound is \cite{toccacelo2025}
\begin{align}
\label{eq:ourlocc-bound}
       F_{c\ell}(t)=\frac{1+\lambda}{1+\lambda+\sin^2(\gamma_\g t)}.
\end{align} 
Here, it is again assumed that Alice draws the initial state from the coherent state ensemble. While challenging, it has been shown that observing violations of such a classical threshold may be possible in the near future with torsion pendulums or harmonically trapped systems \cite{lami2024testing,toccacelo2025}, owing to advances in ground-state cooling of mechanical oscillators.

\subsection*{Critique of the myth: LOCC hypothesis}
There has been considerable recent discussion around what can actually be inferred about quantum gravity from tests that rule out the LOCC hypothesis. Much of this discussion has centered on proposals based on witnessing gravitationally induced entanglement (GIE) \cite{bose2017spin, marletto2017gravitationally}. In the language of the quantum-gravitational imitation game, these proposals challenge the gravitational interaction to generate entanglement between Alice and Bob. Such a task, like the gravitational teleportation dynamics discussed above, is impossible for any classical gravitational interaction constrained to LOCC.

A significant aspect of this discussion is that the LOCC assumption carries substantial ontological baggage, rooted in a particular understanding of locality. The notion of locality assumed in the LO part of LOCC is a notion of \textit{system locality}: it refers to operations that only affect single subsystems of multipartite quantum systems \cite{martinez2023}. This is not necessarily the same as the spatiotemporal \textit{event locality} that underpins field theory, where locality means that interactions are mediated by local field degrees of freedom propagating causally through spacetime. The connection between the two notions becomes apparent only in certain gauge choices---for instance, in the Lorenz gauge, where the dynamics that result in the generation of GIE or gravitational teleportation are a consequence of the mediation by off-shell gravitons \cite{fragkos2022inference}. For different gauge choices, where the observable signatures might no longer be mediated by local fields, rejecting the LOCC hypothesis does not straightforwardly translate into a statement about the quantum nature of the interaction \cite{fragkos2022inference}.

This is precisely why tests based on channel inequalities, such as the classical fidelity bounds discussed above, are theoretically interesting. They enable fine-grained testing of the gravitational interaction. One can envision a hierarchy of inequalities, each corresponding to a different imitation game, in which progressively narrower classes of classical gravitational models are each bounded in a distinct way relative to the quantum null hypothesis. Violating the LOCC bound rules out a wide class of classical models. We can, however, imagine sharper inequalities that target more restricted classical theories, or, conversely, inequalities that bound more general classical models. Such tests would then provide increasingly specific information about the nature of the gravitational interaction.

In his \textit{Conjectures and Refutations}, Karl Popper famously wrote \cite{popper2014conjectures}: "\textit{Science must begin with myths, and with the criticism of myths.}" As our ability to bring increasingly larger systems into the quantum regime advances, the quantum-gravitational imitation games discussed in this essay inch closer to realization. This opens a window to address foundational questions about the nature of gravity and to subject long-standing assumptions to empirical scrutiny. With any luck, this will bring to life new myths worthy of critique. 

\subsubsection*{Acknowledgments}
We thank Alexssandre de Oliveira Junior for help with proofreading the manuscript. We gratefully acknowledge support from the Danish National Research Foundation, Center for Macroscopic Quantum States (bigQ, Grant No. DNRF142), and Novo Nordisk Foundation (Grant No. CBQS NNF 24SA0088433).

\printbibliography 
\end{document}